\documentclass[
  reprint,          
  superscriptaddress,
  amsmath,amssymb,
  aps,pre, longbibliography           
]{revtex4-2}

\usepackage[T1]{fontenc}
\usepackage{graphicx}
\usepackage{microtype}
\usepackage[colorlinks=true,allcolors=blue]{hyperref}

\begin{document}

\title{From spirals to flagellar beating: How pivot-like defects control
semiflexible filament dynamics in motility assays}

\author{Sandip Roy}
\email{mp16001@iisermohali.ac.in}
\affiliation{Indian Institute of Science Education and Research Mohali, Knowledge City,
Sector 81, SAS Nagar – 140306, Punjab, India}
\author{Debasish Chaudhuri}
\email{debc@iopb.res.in}
\affiliation{Institute of Physics, Sachivalaya Marg, Bhubaneswar 751005, India}
\affiliation{Homi Bhabha National Institute, Anushaktinagar, Mumbai 400094, India}
\author{Abhishek Chaudhuri}
\email{abhishek@iisermohali.ac.in}
\affiliation{Indian Institute of Science Education and Research Mohali, Knowledge City,
Sector 81, SAS Nagar – 140306, Punjab, India}

\begin{abstract}
We demonstrate that internal pivot-like defects, arising from rigor mutant motor proteins that bind without stepping, fundamentally reshape the dynamics of semiflexible filaments in two-dimensional motility assays. Using large-scale numerical simulations, we show that such internal pivots establish a previously unrecognized boundary condition, intermediate between free and clamped filaments, that decisively governs filament behavior. Strikingly, by tuning the pivot position, motor activity, and processivity, filaments undergo sharp transitions from tightly wound spiral states to extended, flagella-like beating. Spiral formation is stabilized by a balance between motor-driven forces and bending rigidity, with intermediate stiffness yielding the most robust spirals. Unlike generic active polymer models, our framework isolates the distinct role of rigor-bound motor proteins, revealing how they function as internal control elements governing the transition between spiral and flagellar dynamics. This minimal yet physically grounded model yields experimentally testable predictions and reveals how localized defects can act as key regulators of cytoskeletal organization and dynamics.

\end{abstract}

\maketitle

\section{Introduction}
The cytoskeleton is a dynamic network of semiflexible filaments such as actin and microtubules, together with associated motor proteins~\cite{alberts2015essential, howard2002mechanics, fletcher2010cell}. These components collectively maintain cellular architecture, drive intracellular transport, generate mechanical forces, and facilitate cell motility. Molecular motors such as myosin, kinesin, and dynein convert the chemical energy of ATP hydrolysis into mechanical work, powering directed   motion~\cite{julicher1997modeling, chowdhury2013stochastic, vale2003molecular}. When embedded in networks, motor activity does not simply transport cargo but also imposes active stresses that remodel the cytoskeleton, producing emergent behaviors ranging from contractility and cytoplasmic streaming to large-scale oscillations in cilia and flagella~\cite{ShimmenYokota2004_CytoplasmicStreaming, VicenteManzanares2009_NMMII, riedel2007molecular, sartori2016dynamic, oriola2017nonlinear}.

 To study these dynamics under controlled conditions, in vitro motility assays have been widely employed~\cite{kron1986fluorescent,howard1989movement,vale1994tubulin}, in which surface-bound motors propel filaments that exhibit diverse dynamical states, from collective flows to spiral formation and periodic oscillations~\cite{amos1991bending, bourdieu1995spiral, duke1995gliding, scholey1993motility, uyeda1990myosin, harada1988direct, chaffey2003alberts, sumino2012large, schaller2010polar, ziebert2015microtubules, jiang2014motion, chaudhuri2016forced, shee2021semiflexible, gupta2019morphological, Khalilian2024, karan2024inertia, yadav2024wave}. A central feature of these systems is the strong dependence of filament dynamics on mechanical constraints. Experiments and theory show that pinned ends can induce spirals, while clamped ends promote beating~\cite{bourdieu1995spiral, sekimoto1995symmetry, laskar2015brownian, laskar2013hydrodynamic, jayaraman2012autonomous, chelakkot2014flagellar, isele2016dynamics, vilfan2019flagella}. However, simplifying  motor activity as uniform tangential forcing or noise~\cite{isele2015self,eisenstecken2016conformational,winkler2017active,man2019morphological, winkler2020physics}, overlooks the essential mechanochemical feedback of motor proteins, which both generate forces and respond to them, such as through strain-dependent unbinding that critically shapes filament dynamics.

Of particular interest are rigor mutant motors, which bind strongly to filaments but fail to step due to impaired ATPase activity. Such rigor-bound states effectively create internal pivot-like defects that locally anchor the filament without external clamping~\cite{jon1996crystal,gilbert1995pathway,rayment1993structure,sweeney1998kinetic,rice1999structural,vale1996switches,uyeda1996neck}. These defects are not only biophysically relevant but are also implicated in disorders including hereditary spastic paraplegia, cardiomyopathies, and spinal muscular atrophy~\cite{fink2013hereditary,spudich2014hypertrophic,harms2012mutations}. Yet, despite their clear significance, the influence of internal pivots -- particularly when located away from filament ends -- has not been systematically investigated.

Here, we identify internal pivots as a new class of boundary condition, intermediate between free, pinned, and clamped filaments. Far from being minor perturbations, these defects act as decisive control elements, reorganizing filament dynamics in qualitatively distinct ways. Using large-scale simulations of semiflexible filaments in 2D motility assays, we systematically vary pivot position, motor activity, and processivity to construct a phase diagram, revealing that shifts in pivot location drive transitions from tightly wound spirals to extended, flagella-like beating.  Our analysis -- combining turning number statistics, end-to-end distance fluctuations, tangent-tangent correlations, frequency spectra, and principal component analysis (PCA) of shape dynamics -- reveals how shifting the pivot location reorganizes filament conformations.

In particular, pivots near filament ends robustly stabilize spirals, whereas centrally located pivots induce oscillatory states akin to flagellar beating. These findings establish internal pivot defects as a novel and experimentally accessible mechanism to regulate active filament dynamics.

\begin{figure}[htbp]
\centering
\includegraphics[width=4.25 cm]{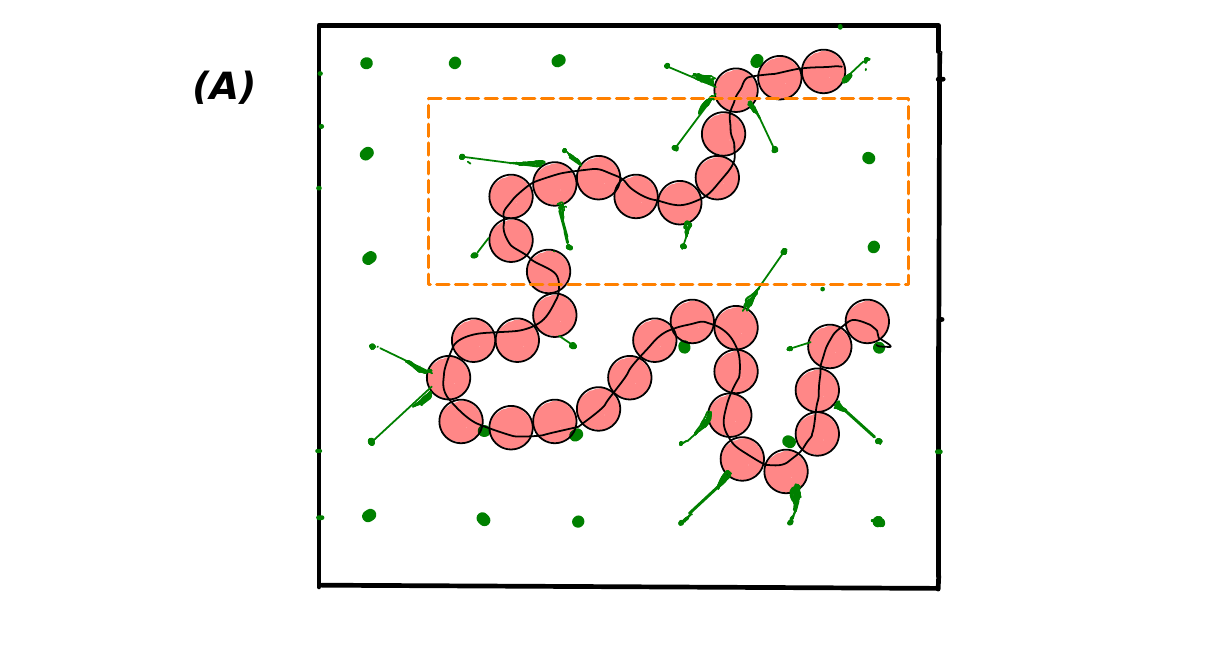}
\includegraphics[width=4.25 cm]{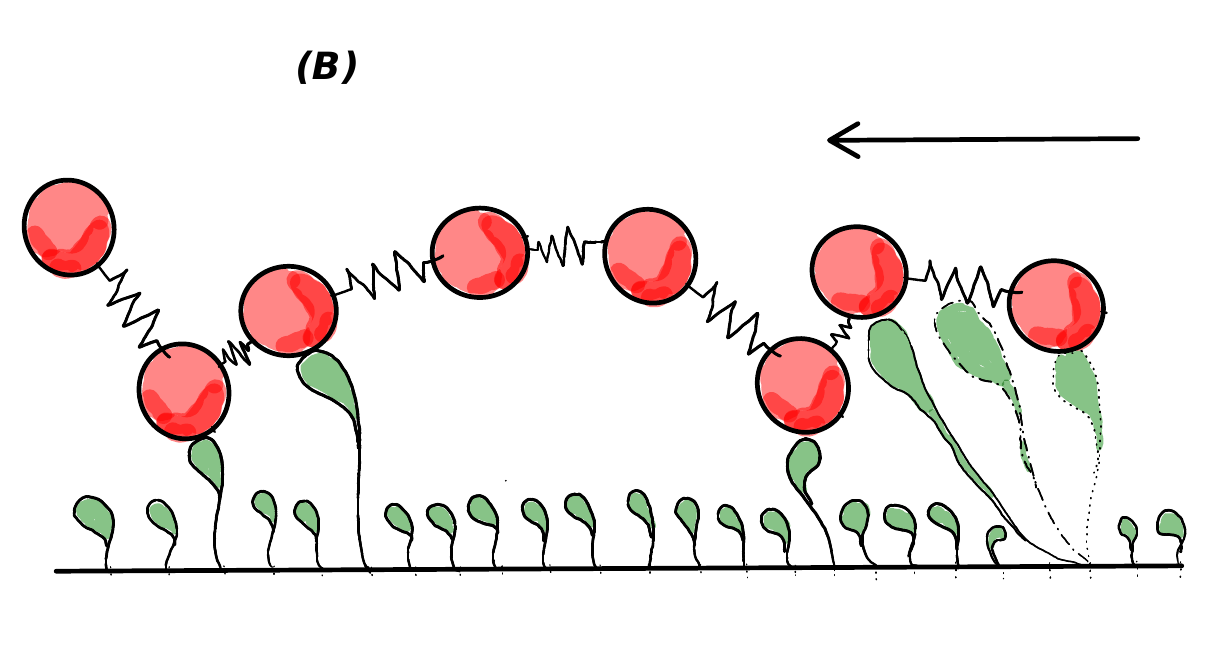}
\caption{
(A) Schematic of the system: a red polymer moves over a grid of green motor proteins (MPs) anchored by their tails. MP heads bind the filament within a capture radius and exert opposing active forces via extensile motion; stalks are modeled as harmonic springs.
(B) Side view (orange box) showing bound MP heads moving along the filament in the indicated direction.
}
\label{fgr:example1}
\end{figure}

\section{Model and Simulation}
The filament is modeled as an extensible semiflexible polymer of $N$ beads with bond vectors ${\bf b}_i = \bf{r}_{i+1} - \bf{r}_i$, with $i = 1, 2, ...., N-1$ and ${\bf r}_1, {\bf r}_2, ..., {\bf r}_N$ denote the monomer positions. The stretching energy cost is given by
\begin{equation}
 {\cal E}_s = \sum_{i=1}^{N-1}\frac{A}{2r_0}|{\bf b}_i - r_0\hat{t}_i|^2,
 \end{equation}
 where $\hat{t}_i = {\bf b}_i/|{\bf b}_i|$ denotes the local tangents, $A$ is the bond stiffness and $r_0$ is the equilibrium bond-length. The total contour length of the polymer is $L = (N-1)r_0$. The bending-energy cost is expressed as
 \begin{equation}
 {\cal E}_b = \sum_{i=1}^{N-2}\frac{\kappa}{2r_0}|\hat{t}_{i+1}-\hat{t}_i|^2,
 \end{equation}
where $\kappa$ denotes the bending rigidity. For an equilibrium worm-like chain, the dimensionless stiffness parameter $u = L/l_p$ provides a measure of the semiflexibility of the chain where $l_p = 2\kappa/(d-1)k_BT$ is the persistence length in $d$-dimensions. Such a chain transitions between Gaussian chain (at $u \approx 10$) to rigid-rod (at $u \approx 1$) with coexistence around $3\lesssim u \lesssim 4$~\cite{dhar2002triple, Chaudhuri2007}.  In our simulations of $N=64$ bead chain, we choose ${\kappa}/{\sigma k_BT} = 9.46$, corresponding to $u \approx 3.33$. 

To enforce self-avoidance, we include the Weeks-Chandler-Anderson (WCA) interaction, the repulsive part of Lennard-Jones potential between non-bonded beads, using an energy cost 
\begin{equation}   
  {\cal E}_{WCA}  = 
\begin{cases}
    4\epsilon\, [(\frac{\sigma}{r})^{12} - (\frac{\sigma}{r})^6 + \frac{1}{4}] ,& \text{if } r< 2^\frac{1}{6}\sigma \\
    0,              & \text{otherwise}
\end{cases}
\end{equation}
where $\epsilon$ and $\sigma$ set the energy and length scales, respectively. The total energy cost is ${\cal E} = {\cal E}_s + {\cal E}_b + {\cal E}_{WCA}$. We use the equilibrium bond-length $r_0 = 1.0\, \sigma$.

In the motility assay (Fig.\ref{fgr:example1}), a polymer lies on a substrate densely covered with immobilized motor proteins (MPs) that are modeled as linear springs with stiffness $k_m$. Each MP is pivoted by its tail at fixed positions on a 2D square lattice with uniform density $\rho$. 
While MP heads can attach to the conjugate filament at a rate $\omega_{\text{on}}$ through a diffusion-limited process, provided a polymer bead lies within the capture radius $r_c$.

Once attached, motor protein (MP) heads move along the filament, as described in detail later. Extensions from their equilibrium length, denoted by $\Delta \mathbf{r}$, generate a load force on the motors given by
$\mathbf{f}_l = -k_m \Delta \mathbf{r}$.
The detachment rate of MPs increases exponentially with this load,
\begin{equation}
\omega_{\text{off}} = \omega_0 \exp\left(\frac{f_l}{f_d}\right),
\end{equation}
where $f_l=|\mathbf{f}_l |$, $\omega_0$ is the zero-load (intrinsic) detachment rate, and $f_d$ sets the characteristic force scale for detachment. Such a choice of attachment and detachment rates introduces an asymmetry that violates detailed balance.

The extension $\Delta \mathbf{r}$ of attached MPs evolves dynamically through both passive motion of the polymer and active translocation of the motor head along the filament backbone. During translocation, the head may occupy positions along the bonds between polymer beads, and the resulting elastic force $\mathbf{f}_l$ is distributed to adjacent beads via a lever rule weighted by the motor’s proximity to each bead.

The motor’s active speed along the filament is also load-dependent and is modeled as:
\begin{equation}
v_a(f_t) = \frac{v_0}{1 + d_0 \exp(f_t / f_s)},
\end{equation}
where $v_0$ is the bare speed without load, $f_t = \mathbf{f}_l \cdot \hat{t}$ is the tangential component of the load force, $d_0 = 0.01$~\cite{chaudhuri2016forced}, and $f_s$ denotes the stall force. Throughout this study, the unidirectional movement of the motor protein along the filament is assumed to proceed in the direction from bead   $i=1$ to $i=N$. As a result, the filament experiences a tangential active force acting in the opposite dirtection.

Simulations combine stochastic motor protein dynamics with a velocity-Verlet algorithm for the chain coupled to a Langevin heat bath characterized by isotropic friction $\gamma$ per bead and constant temperature $k_B T = 1.0 \varepsilon$. 
To minimize bond length fluctuations, the bond stiffness is set to $A = 100 \varepsilon / \sigma$. We use a capture radius $r_c = 0.5 \sigma$~\footnote{Noting the small size of MPs compared to the polymer.} and the MP density to $\rho = 4 \sigma^{-2}$. The spring constant for the motor stalk is set as $k_m = A / \sigma$. Active forces exceed thermal fluctuations, with $f_s = 2 k_B T / \sigma$ and $f_d = f_s$.

The polymer–motor protein dynamics are characterized by the dimensionless Péclet number 
$\text{Pe} = \frac{v_0 L^2}{D r_0}$
and the bare processivity
$\Omega = \frac{\omega_{\text{on}}}{\omega_{\text{off}} + \omega_0}$. 
The characteristic equilibrium diffusion time for the polymer to traverse its contour length is
$\tau = \frac{L^3 \gamma}{4 \sigma k_B T}$,
which sets the simulation time unit. For numerical stability, we use a time step
$dt = 1.6 \times 10^{-8} \, \tau$,
and run simulations for $2 \times 10^9$ steps, discarding the first $10^9$ steps to reach steady state. A summary of all simulation parameters is provided in Table~\ref{table}.

In the following, we refer to the imposed constraint -- implemented in simulations by fixing the lateral position of a monomer -- as a pivot defect. For clarity, we use the term {\it end pivot} when the defect is located at a filament end, and {\it interior pivot} when it is positioned away from the ends.

\begin{table}[h!]
\small
  \caption{Different parameters and their numerical values used in the simulation: 
  }
  \label{tbl:example1}
  \begin{tabular*}{0.48\textwidth}{@{\extracolsep{\fill}}lll}
    \hline
    Parameters & Definition & Values \\
    \hline
    N & Number of polymer beads & 64 \\
    $r_0$ & Bond length  & 1.0~$\sigma$\\
    $\gamma$ & Isotropic friction per bead & 1\\
    $k_BT$ & Energy scale  & 1\\
    $r_c$ & Capture radius & $0.5~\sigma$ \\
    $A$ & Bond stiffness &100 $k_BT/\sigma$ \\
    $\rho$ &Density of MP &$3.8~\sigma^{-2}$\\
    $f_d$ & Detachment force &2 $k_BT/\sigma$\\
    $f_s$ & Stall force &2 $k_BT/\sigma$\\
    $d_0$ & Force sensitivity parameter &0.012\\
    $\gamma_{m}$ & Frictional coeffecient of MP &$0.1~\gamma$\\
    $k_m$ & Elastic coeffecient of MP &100 $k_BT/\sigma^2$ \\  
    $\text{Pe}$ &P\'eclet number $(v_0L^2/Dr_0)$ & $1.9\times10^4$\\
      & & -- $ 29.7\times10^4$\\
    $\Omega$ & Bare processivity $\left(\frac{\omega_{\text{on}}}{(\omega_{\text{off}} + \omega_0)}\right)$ & 0-1\\
    \hline
  \end{tabular*}
  \label{table}
\end{table}

\section{Results}
We characterize pivot-induced transitions using the following complementary observables: turning number statistics (including kurtosis) to quantify spiral formation, and principal component analysis (PCA) frequencies to capture the onset of flagellar beating.

\subsection{Morphology}
We begin by analyzing the morphology of the polymer with a single pivot defect. The defect location is parameterized by $m/N$, where $m$ denotes the index of the monomer at which the pivot constraint is imposed along a polymer of $N$ beads. For instance, $m = 1$ places the pivot defect at the leading end of the polymer, while increasing $m$ ($m = 2, 3, \ldots$) progressively shifts the defect inward along the filament contour. Representative configurations are shown in Fig.~\ref{morph}.

\subsubsection{Morphologies with end pivot}

\begin{figure*}[htb!]
\centering
\includegraphics[width=\linewidth]{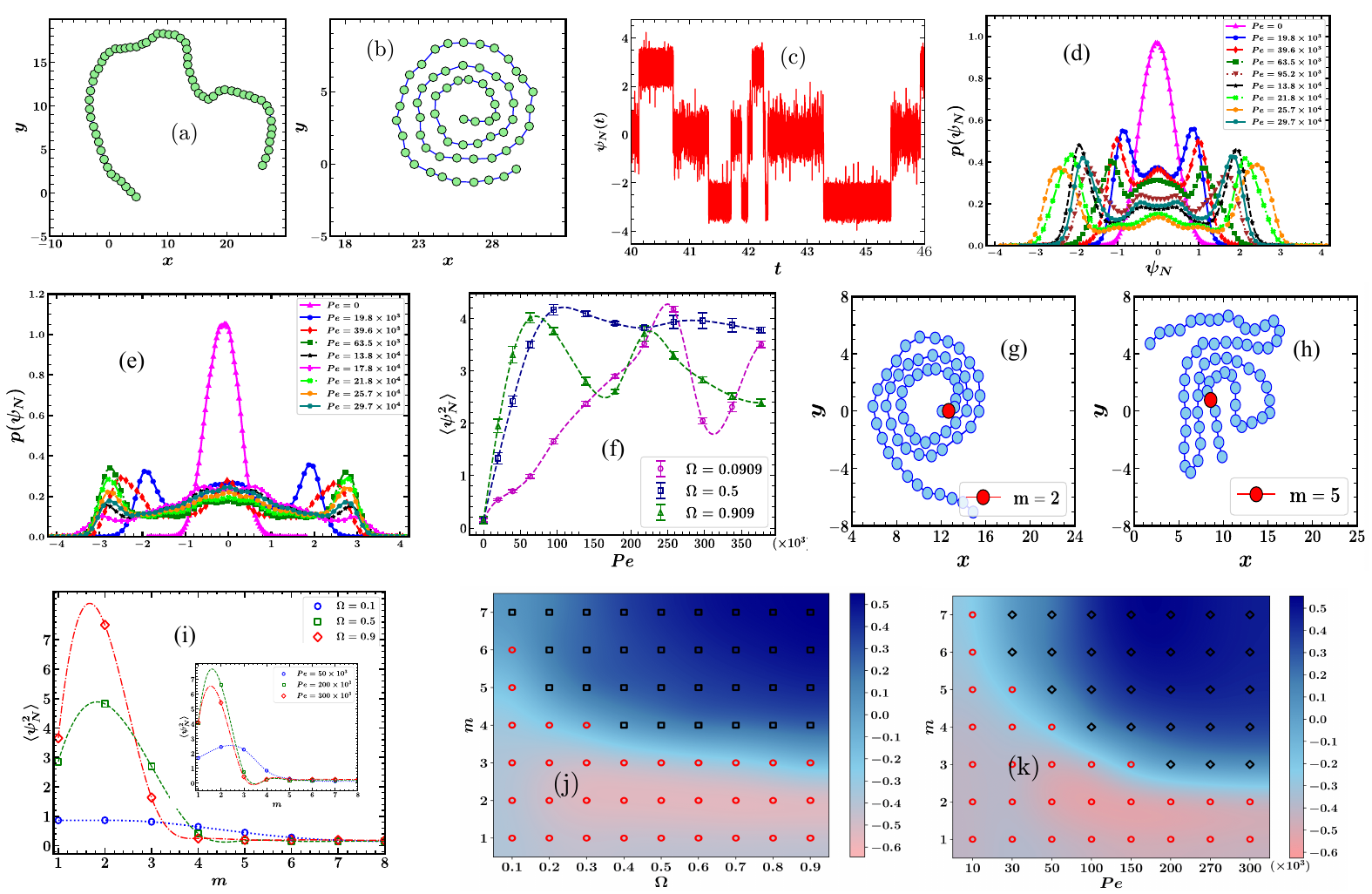}
\caption{
A semiflexible polymer chain of length $L = 63\sigma$ with persistence ratio $u = 3.33$ on a motility assay. 
(a) Typical polymer conformation at $\text{Pe} = 0$. 
(b) Conformations with a single point defect at the leading end at $\text{Pe} = 9.5 \times 10^4$. 
(c) Time series of the turning number $\psi_N(t)$ for a free polymer at $\text{Pe} = 9.5 \times 10^4$ and $\Omega = 0.5$.  
(d) Steady-state turning number distribution $p(\psi_N)$ for $\Omega = 0.1$ and (e) for $\Omega = 0.9$ at various $\text{Pe}$ values as indicated.  
(f) Mean squared turning number $\langle \psi_N^2 \rangle$ versus $\text{Pe}$ for $\Omega = 0.09$, $0.5$, and $0.9$.  
(g)–(h) Representative conformations for defect locations $m = 2$~(g) and $m = 5$~(h).
(i) $\langle \psi_N^2 \rangle$ versus defect location $m$ for $\Omega = 0.1$, $0.5$, and $0.9$ at fixed $\text{Pe} = 9.92 \times 10^4$. Inset: variation with $\text{Pe}$ at $\Omega = 0.5$. 
(j)–(k) Phase diagrams in the $(m, \Omega)$ plane at fixed $\text{Pe} = 9.92 \times 10^4$~(j) and in the $(m, Pe)$ plane at fixed $\Omega = 0.5$~(k), with color maps representing excess kurtosis $\mathcal{K}$.  
}

\label{morph}
\end{figure*}

At zero activity ($\text{Pe} = 0$), the semiflexible polymer with significant bending rigidity adopts open-chain conformations (Fig.~\ref{morph}(a)). Introducing a single pivot defect at the leading end $i=1$ of the filaments' active motion induces buckling and spontaneous spiral formation for any non-zero $\text{Pe}$ and  $\Omega$ values (Fig.~\ref{morph}(b)\,). These spirals rotate either clockwise or counterclockwise, with the number of turns increasing with $\text{Pe}$, while preserving chiral symmetry. To quantify spiral formation, we use the turning number $\psi_N$, defined as the cumulative angle between successive bond vectors across the filament: $\psi_N = 0$ for a straight chain, $\psi_N = \pm 1$ for a full loop, and larger $|\psi_N|$ values indicate tighter spirals with a larger number of turns. In two-dimensions it can be defined as $\psi_N=(1/2\pi)\sum_{i=1}^N [\theta_{i+1} - \theta_i]$ where $\theta_i$ denotes the angle subtended by the $i$-th bond with $x$-axis.  The time evolution of $\psi_N$ for a freely moving filament on the motor protein assay, in the absence of pivot defects, is shown in Fig.~\ref{morph}(c). The data correspond to $\text{Pe} = 9.5 \times 10^4$ and $\Omega = 0.5$.

\vskip 0.2cm
\noindent
{\em Turning number distribution:}
The time series $\psi_N(t)$ reveals that the filament morphology undergoes smooth transitions between right- and left-handed spiral states via intermediate open-chain configurations ($\psi_N = 0$), reflecting the coexistence of open filaments and spirals of both chiralities.  Previous studies have investigated the morphological transitions from predominantly open chains to stable spirals occurring beyond a critical activity in freely moving filaments on motor protein assays and active filaments~\cite{shee2021semiflexible,karan2024inertia}. 
Introducing a pivot defect at the leading end, $m=1$, fundamentally alters filament behavior by promoting spiral formation at infinitesimally small activity, effectively shifting the transition point to vanishing $\text{Pe}$ and $\Omega$. Three key observations characterize this effect:

(i) Symmetry dictates that the average value of $\psi_N$ remains close to zero, and the distribution $p(\psi_N)$ is symmetric and trimodal, exhibiting peaks near $\psi_N \approx \pm 1$ (at small $\text{Pe}$) and $\psi_N \approx 0$ (see Fig.~\ref{morph}(d)). 

(ii) Even at low processivity ($\Omega = 0.1$) and small $\text{Pe}$,  $p(\psi_N)$ is dominated by side peaks indicating the coexistence of stable spirals and a metastable open state (Fig.~\ref{morph}(d)). As $\text{Pe}$ increases, spiral peaks move outward, corresponding to tighter spirals, but at very high $\text{Pe}$, a re-entrant shift to smaller $|\psi_N|$ occurs due to increased motor detachment, weakening filament activity and spiral stability.

(iii) At higher processivity ($\Omega = 0.9$), the side peaks occur at larger $\psi_N$ values, signaling tighter but less stable spirals, as evidenced by reduced peak heights (Fig.~\ref{morph}(e)). Increasing $\text{Pe}$ at this higher $\Omega$ similarly triggers a re-entrant transition toward smaller $|\psi_N|$. 

\noindent
These results highlight that spiral stability depends sensitively on the interplay between motor activity and attachment-detachment dynamics. 

The initial tightening of spirals with increasing activity can be captured by the decrease in the end-to-end distance $R$. A mean-field torque balance predicts that $R$ scales as $({\rm Pe}\, \Omega)^{-1/3}$, where the active torque scales with $\text{Pe} \, \Omega$ and is balanced by the filament's bending rigidity.
As detailed in Appendix~\ref{appendix:spiralsize}, our simulation data shows reasonable agreement with this scaling form over an intermediate range of activity.

\vskip 0.2cm
\noindent
{\em Lower order moments of turning number:}
We analyze the first two steady-state moments of the turning number $\psi_N$, as they are more easily accessible in experiments than the full distribution. Due to chiral symmetry ($p(\psi_N) = p(-\psi_N)$), the mean $\langle \psi_N \rangle = 0$, and all odd moments vanish. The mean square fluctuation, $\langle \psi_N^2 \rangle$, quantifies the effective turning, with higher values corresponding to better stability and larger turning number of the spirals.

In Fig.~\ref{morph}(f), we present the variation of the second moment, $\langle \psi_N^2 \rangle$, as a function of $\mathrm{Pe}$ for different values of $\Omega$ at fixed $m = 1$. This moment is determined by both the stability (i.e., probability) of the spiral states and their turning numbers.

For low $\Omega$ ($=0.1$),  $\langle \psi_N^2 \rangle$ initially increases almost linearly with $\mathrm{Pe}$. At higher $\mathrm{Pe}$, we observe a periodic fluctuation that corresponds to the lateral shifts of secondary peaks in the distribution. At intermediate $\Omega = 0.5$, the increase with $\mathrm{Pe}$ is sharper, though the amplitude of the fluctuation at larger $\mathrm{Pe}$ becomes less significant. For high $\Omega$ values ($=0.9$), large fluctuations reappear with increasing $\mathrm{Pe}$.

These periodic fluctuations arise because the relative contributions of the secondary peaks at large $|\psi_N|$ change substantially compared to the central peak near $\psi_N \approx 0$. Since $\langle \psi_N^2 \rangle$ is particularly sensitive to the probability weight at high turning numbers, variations in the relative strength of these peaks cause the observed fluctuations in the second moment.

\subsubsection{Morphologies for a Pivot at Subsequent Positions along the Polymer}

Shifting the pivot defect from the leading end of the polymer to internal positions results in significant changes in morphology. With the pivot positioned near the end ($m = 2$), the filament still forms spirals as the free end rotates around the pivot point (Fig.~\ref{morph}(g)). However, for larger $m$ (e.g., $m = 5$), spiral formation is suppressed (Fig.~\ref{morph}(h)). The short, rigid segment between the head and pivot remains mostly stationary and straight, as geometric constraints and bending rigidity prevent it from deforming despite motor activity. The downstream segment remains dynamic but adopts open, rather than coiled,  configurations. At high $m$ and activity levels, the filament exhibits flagella-like beating motions.


\vskip 0.2cm
\noindent
{\em Lower order moments of turning number:} Figure~\ref{morph}(i) shows the variation of $\langle \psi_N^2 \rangle$ with $m$ for different $\mathrm{Pe}$ values at fixed $\Omega$, illustrating the strong influence of activity and processivity on spiral morphology. At low processivity ($\Omega = 0.1$), changes in pivot position have little impact, and the filament remains mostly in an open-chain state. With higher $\Omega$, enhanced motor attachment facilitates the transfer of active forces, leading to stable spiral formation for $m \leq 2$. For larger $m$, the filament predominantly adopts open configurations with occasional metastable spirals. Notably, the most stable spirals occur at $m = 2$. A similar pattern emerges when varying $\mathrm{Pe}$ at high $\Omega$ (inset, Figure~\ref{morph}(i)\,).

{
We note that the apparent enhancement of spiral stability at $m=2$ compared to $m=1$ partly reflects the discrete bead--spring representation of the polymer. In simulations with finer discretization, this difference is reduced, and the behavior of near-end pivots becomes essentially equivalent. 
We confirm this through simulations with finer discretization (more beads, same total length). As shown in Fig.~\ref{discrete} in Appendix~\ref{appendix:discretization}, the difference in spiral formation probability (with $\psi_N \approx 3$) between $m=1$ and $m=2$ decreases with finer discretization [Fig.~\ref{discrete}(a)] compared to the coarser case [Fig.~\ref{discrete}(b)]. This suggests that in the continuum limit, pivoting at infinitesimally separated points yields negligible differences -- though finite shifts still produce distinct morphology and dynamics. Importantly, the spiral--beating transition and the phase boundaries remain qualitatively robust.
}

{
\subsubsection{Spiral phase transition}
\label{sec:phase_diagram}
We summarise the dependence of filament state on pivot location, activity, and processivity using the excess kurtosis ${\cal K} = \langle\psi_N^4\rangle/3\langle\psi_N^2\rangle^2 - 1$ of the turning number distribution $p(\psi_N)$ as the order parameter (Figs.~\ref{morph}(j,k)). By design, ${\cal K}$ is more sensitive to the nature of the probability distribution than the second moment. For the multimodal distribution of $\psi_N$ associated with spiral states, negative ${\cal K}$ values indicate spiral dominance. A pure spiral state yields ${\cal K} = -2/3$, while coexistence with open chains gives ${\cal K} = -1/3$~\cite{karan2024inertia}. In the absence of spirals, ${\cal K} = 0$, corresponding to a Gaussian distribution centered at $\psi_N = 0$. However, conformational fluctuations toward spiral states can enhance tail weights of such unimodal distributions, making ${\cal K}$ positive.

The phase diagrams in the $(m, \Omega)$ and $(m, Pe)$ planes (Figs.~\ref{morph}(j,k)) reveal that for near-end pivots ($m \leq 2$), even moderate activity yields negative $-2/3 < {\cal K} < -1/3$ identifying robust spirals. As the pivot moves inward, ${\cal K}$ increases, signalling a suppression of spirals and a crossover to open chains which, as we show later, display extended beating states. At lower activity (smaller $\text{Pe}$ or $\Omega$), the crossover occurs via inward shift of pivot positions, marked by increasing $m$, as shown in the phase diagrams Figs.~\ref{morph}(j,k)  (also see Appendix~\ref{appendix:kurtosis}).
}

\begin{figure}[htbp]
\centering
\includegraphics[width=8 cm]{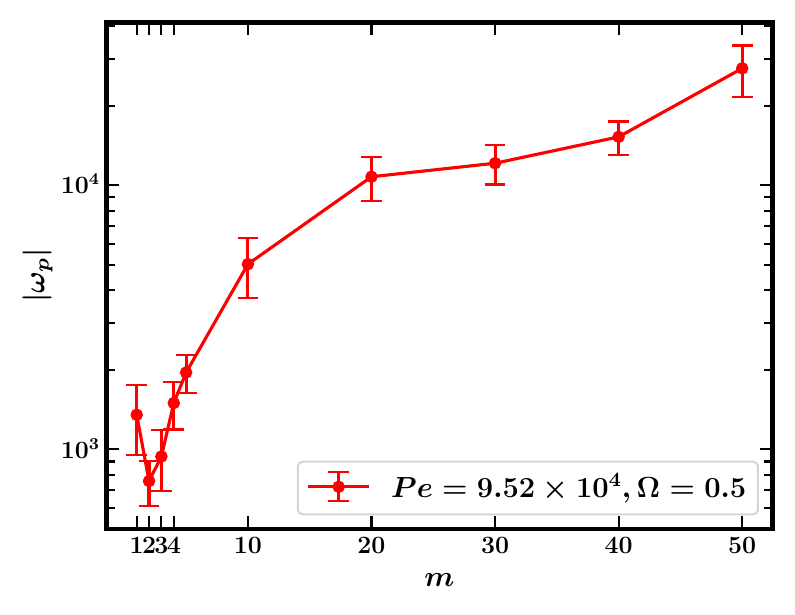}
\caption{Limit cycle frequency from PCA analysis as a function of pivot location $m$ for fixed $\Omega = 0.5$ and $\text{Pe} = 9.52\times10^4$. Frequencies are expressed in units of the inverse characteristic time $\tau^{-1}$.}
\label{fig:transition}
\end{figure}

\begin{figure*}[htbp]
\centering
\includegraphics[width=\linewidth]{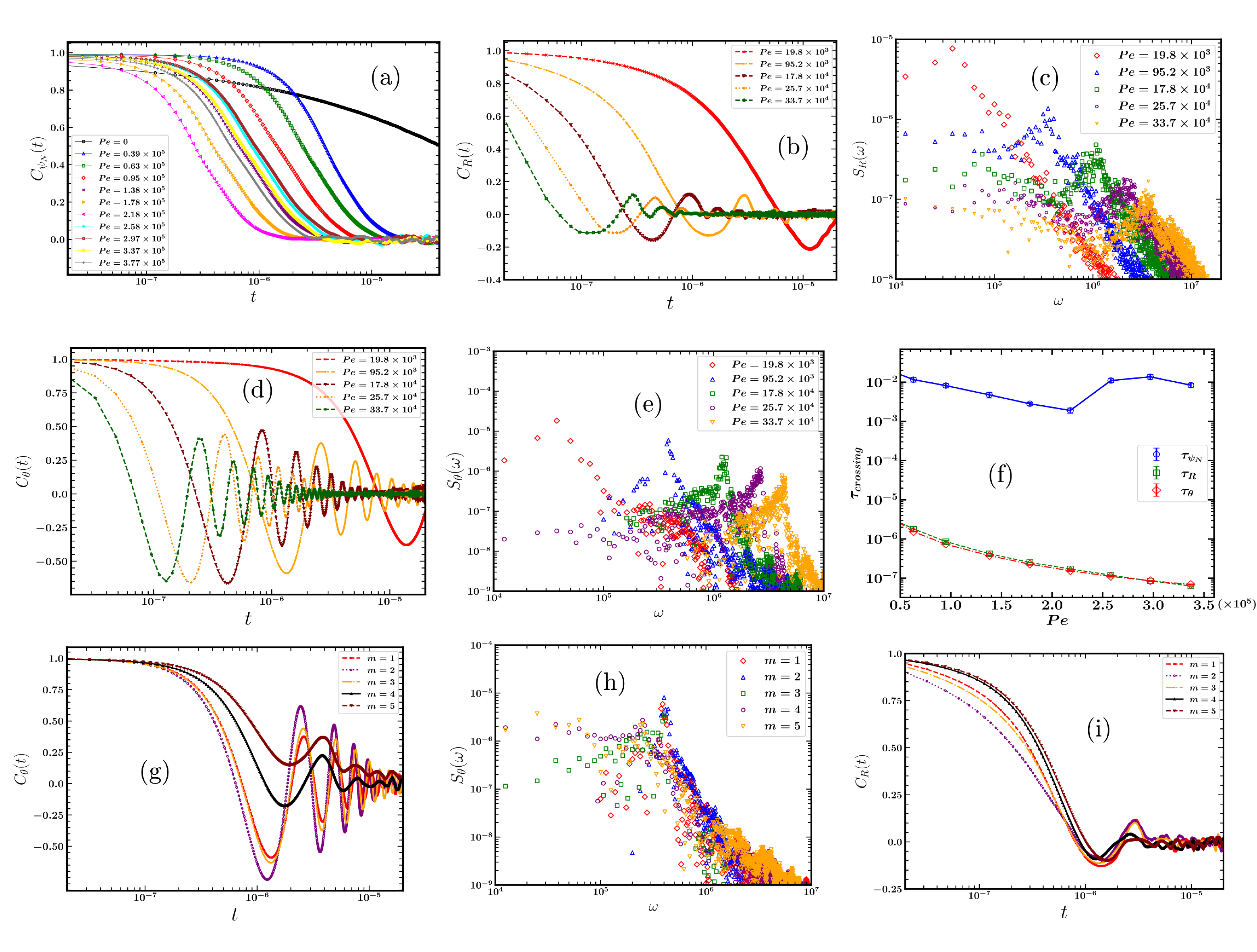}
\caption{Dynamics of a semiflexible polymer chain of length $L = 63\sigma$ with persistence ratio $u = 3.33$ on a motility assay.
(a) Autocorrelation of $C_{\psi_N}(t)$ as a function of time for $\Omega = 0.5$, with time measured in units of $\tau$.
(b) Autocorrelation of the end-to-end distance $C_R(t)$ for five values of Pe at fixed processivity $\Omega = 0.5$.
(c) Power spectral density $S_R(\omega)$ of the end-to-end separation, with frequency $\omega$ in units of $1/\tau$.
(d) Autocorrelation of the end-to-end vector angle $C_\theta(t)$ for varying Pe.
(e) Power spectral density corresponding to panel (d), with $\Omega = 0.5$.
(f) First zero-crossing times of $C_{\psi_N}(t)$, $C_R(t)$, and $C_\theta(t)$ as a function of Pe.
(g) Angular autocorrelation $C_\theta(t)$ for five defect locations ($m=1,2,3,4,5$) at $\Omega = 0.5$.
(h) Power spectral density corresponding to panel (g).
(i) Autocorrelation of the end-to-end separation $C_R(t)$ for the same defect locations as in panel (g).}
\label{dynamics}
\end{figure*}

\vskip 0.2cm
\noindent
\subsubsection{Transition from spiral to flagella}
For intermediate pivot positions, we quantified the transition from spiral to flagellar motion using principal component analysis (PCA) of the filament's shape dynamics (see Appendix~\ref{appendix:pca} for details)~\cite{ma2014active}. The tangent angles along the filament were decomposed into dominant shape modes, and the trajectory in the space of the first two mode amplitudes, $(B_1, B_2)$, was analyzed. During flagella-like beating, the parametric plot of these modes forms a closed limit cycle,  although stochastic, reflecting periodic motion. From the phase progression along this cycle, we extracted a mean beating frequency. This frequency reaches a minimum at pivot position $m = 2$, corresponding to a tight spiral (Fig.~\ref{fig:transition}), then increases with $m$ as the filament transitions to an open configuration, eventually stabilizing for pivots far from the filament's leading end, capturing the characteristic frequency of flagellar beating.

\subsection{Dynamics}
Having characterized the polymer morphologies and transitions, we now examine filament dynamics with a single pivot defect using two-time autocorrelation functions of the turning number and end-to-end vector.

\subsubsection{Filament dynamics with terminal pivot}
\vskip 0.2cm
\noindent
{\em Correlation in turning number:} 
Figure~\ref{dynamics}(a) shows the two-time autocorrelation of the turning number, $C_{\psi_N}(t) = \langle \psi_N(t)\psi_N(0) \rangle / \langle \psi_N^2(0) \rangle$, for various $\text{Pe}$ at fixed  $\Omega=0.5$. At $\text{Pe} = 0$, the polymer remains in an open-chain state, with $C_{\psi_N}(t)$ displaying a single exponential decay, characteristic of a stochastic relaxation within the open state. For a single end pivot, the $\text{Pe}$ values shown in the figure stabilize spiral states, leading to a double-exponential decay. The second exponential with slower decay reflects dynamical transformations between open and spiral morphologies, similar to the situation in the absence of defects, though with a significantly shallower crossover as the pivot renders the switching between the states easier. The relaxation time $\tau_{\psi_N}$ initially decreases with increasing $\mathrm{Pe}$ as activity promotes transitions, but rises at higher $\mathrm{Pe}$ because enhanced motor detachment reduces effective filament driving; see Figure~\ref{dynamics}(f).

\vskip 0.2cm
\noindent
{\em Dynamics of the end-to-end vector:} Figure~\ref{dynamics}(b) presents the autocorrelation of the end-to-end separation, with the end-to-end vector defined as ${\bf R}(t) = R(t)\exp[i\theta(t)]$, and the correlation function given by $C_R(t) = \langle \delta R(t) \delta R(0) \rangle/ \langle \delta R^2(0) \rangle$ where $\delta R(t)=R(t)-\langle R(t) \rangle$. With increasing $\text{Pe}$, $C_R(t)$ decays faster and exhibits oscillations at longer times, reflecting the stretching and recoiling dynamics of spirals. The relaxation time $\tau_R$ decreases monotonically with $\text{Pe}$ (Fig.~\ref{dynamics}(f)). This timescale produces a peak in the spectral density, which shifts to higher frequencies with increasing $\text{Pe}$ (Fig.~\ref{dynamics}(c)).

The end-to-end distance distributions further illustrate this trend: 
as $\text{Pe}$ increases, the peak of $p(R)$ shifts to lower values, 
indicating tighter coiling of the filament. 
These distributions, shown in Fig.~\ref{fig:pend} in Appendix~\ref{appendix:endtoend}, complement the time-resolved 
autocorrelation functions.

The end-to-end orientation autocorrelation, $C_\theta(t) = \langle \cos(\theta(t) - \theta(0)\,) \rangle$, shows a similar trend (Fig.~\ref{dynamics}(d)). It decays exponentially with superimposed oscillations whose frequency increases and amplitude decreases with $\text{Pe}$. These oscillations arise from spiral rotation about the pivot. The corresponding relaxation time $\tau_\theta$ also decreases with increasing activity (Fig.~\ref{dynamics}(f)), and has similar values as the size relaxation time $\tau_R$. The characteristic frequency of the oscillation appears as a peak in the angular spectral density, which shifts to higher frequencies with $\text{Pe}$ (Fig.~\ref{dynamics}(e)).

\subsubsection{Filament dynamics with intermediate pivot}
As previously noted, shifting the defect away from the tail induces a transition from spiral to flagella-like beating conformations, which is also reflected in the dynamics of the end-to-end separation $\mathbf{R}(t)$ (Figs.~\ref{dynamics}(g-i)). The directional autocorrelation $C_\theta(t)$ reveals distinct decay timescales and persistent oscillations across all pivot positions $m$. For $m = 2, 3, 4$, these oscillations arise from spiral rotation around the pivot, with their amplitude decreasing as $m$ increases, indicating suppression of spiral motion. In contrast, at larger $m$ ($m = 5$), oscillation amplitude grows significantly along with an increased frequency, signaling a transition to flagella-like beating in the free filament segment. Figure~\ref{dynamics}(h) presents the corresponding power spectral density, while Fig.~\ref{dynamics}(i) shows the correlation of the end-to-end distance.

\subsection{Importance of bending rigidity in sustaining spirals}
\begin{figure}[t]
\centering
\includegraphics[width=\columnwidth]{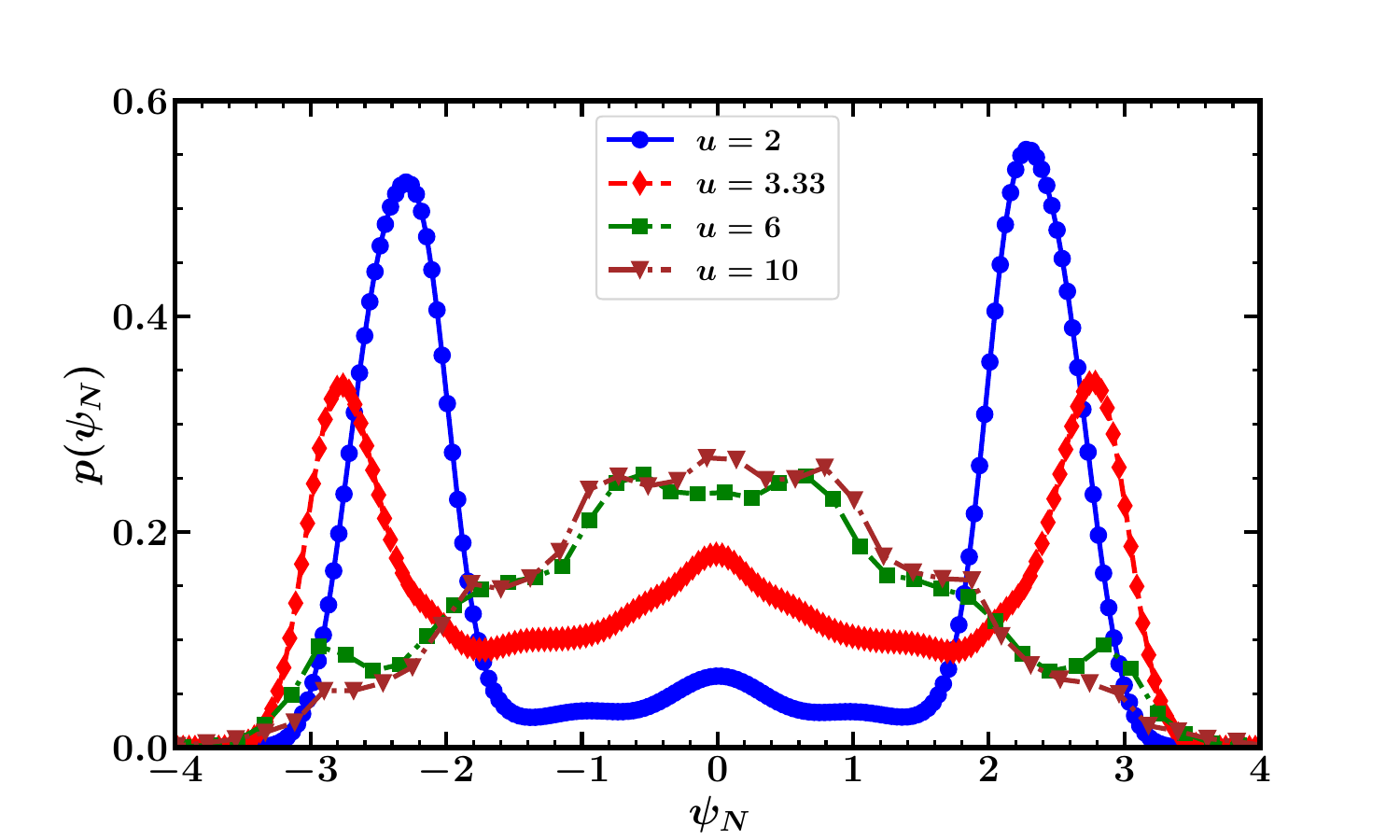}
\caption{Probability distribution of the turning number $p(\psi_N)$ for a polymer of length $ L = 63\sigma$, with a processivity $ \Omega = 0.5$. Results are shown for polymers with varying stiffness, characterized by the stiffness parameter $u$, highlighting the transition from stiff to flexible regimes. }
\label{vary_persistence}
\end{figure}

To assess the role of bending rigidity in spiral formation, we varied the persistence ratio $u$ from stiff ($u \approx 1$) to flexible ($u \approx 10$), with the leading end of the polymer pivoted. Keeping $\Omega$ and $\text{Pe}$ fixed at intermediate values, we analyze the turning number distribution $p(\psi_N)$ (Fig.~\ref{vary_persistence}). For stiff filaments ($u = 2$), stable spirals form with fewer turns ($\psi_N \approx \pm 2$). At a smaller rigidity ($u = 3.33$), spirals with higher turning numbers ($\psi_N \approx \pm 3$) emerge, though with less probability. For flexible chains ($u = 6,\,10$), spiral stability is lost, and the filament primarily adopts open-chain or unstable coiled states. The preference for intermediate stiffness arises from a balance between bending modulus and active torque. If the filament is too stiff, active forces are insufficient to bend it into a spiral. Conversely, if it is too flexible, the filament cannot sustain the curvature required for a stable spiral, leading instead to transient coiled states interspersed with dominantly open conformations. We have verified that this effect persists under a finer discretization of the polymer.

\section{Conclusion}

We investigated the dynamics of active semiflexible filaments in two-dimensional motility assays in the presence of internal pivot-like defects, motivated by rigor-bound motors that bind strongly but fail to step~\cite{jon1996crystal,gilbert1995pathway,rayment1993structure,sweeney1998kinetic,rice1999structural,vale1996switches,uyeda1996neck} and are implicated in diverse disease phenotypes~\cite{fink2013hereditary,spudich2014hypertrophic,harms2012mutations}. Our results demonstrate that these defects constitute a previously unrecognized class of boundary condition, intermediate between the well-studied free, pinned, and clamped limits. Far from being minor perturbations, internal pivots decisively reorganize filament dynamics by introducing localized anchoring at arbitrary positions along the contour. 

A central finding of this study is that shifting the pivot location qualitatively alters filament dynamics, driving a transition from tightly coiled spiral states to extended, flagella-like beating. Pivots near filament ends stabilize spirals with characteristic bimodal turning-number distributions, while centrally placed pivots favor oscillatory conformations and flagella-like periodic beating. This pivot-induced control of filament organization is further modulated by motor activity and processivity, highlighting the cooperative interplay between active driving, filament elasticity, and internal constraints.

We also show that intermediate filament stiffness promotes the most robust spiral states, reflecting a balance between active torque and bending rigidity. These findings suggest that internal pivots are not only a minimal physical analogue for rigor-bound motors but also an experimentally accessible mechanism to regulate filament organization in motility assays.

In summary, this work identifies internal pivot-like defects as key control elements of active filament dynamics, showing how localized anchoring constraints can reorganize motion between spiral and flagellar states through the simple tuning of pivot position. This mechanism provides testable predictions for in vitro molecular motor assays, offers design strategies for steering oscillations and pattern formation in synthetic active matter, and carries broader implications for cell biology, where anchoring proteins, cross-links, or structural defects may employ similar principles to regulate cytoskeletal function.

\section*{Author Contributions}
Sandip Roy: Investigation; Software; Data curation; Formal analysis; Validation. 
Debasish Chaudhuri: Conceptualization; Supervision; Methodology; Writing -- original draft, review \& editing. 
Abhishek Chaudhuri: Conceptualization; Supervision; Methodology; Writing -- original draft, review \& editing.

\section*{Conflicts of interest}
There are no conflicts to declare.

\section*{Data availability}
All data generated or analysed during this study are included in the article. 

\section*{Code availability}
The code underlying this study is available from the corresponding author upon reasonable request by qualified researchers.

\section*{Acknowledgements}
D.C. acknowledges support from DAE (India) under grant 1603/2/2020/IoP/R\&D-II/150288, a Visiting Professorship from C.Y. Cergy Paris Universit{\'e}, and an Associateship at ICTS-TIFR, Bangalore. AC acknowledges support from the Indo-German grant (IC-12025(22)/1/2023-ICD-DBT).



\appendix

\section{Activity-dependence of spiral size}
\label{appendix:spiralsize}
\begin{figure}[t!]
\centering
\includegraphics[width=0.9\columnwidth]{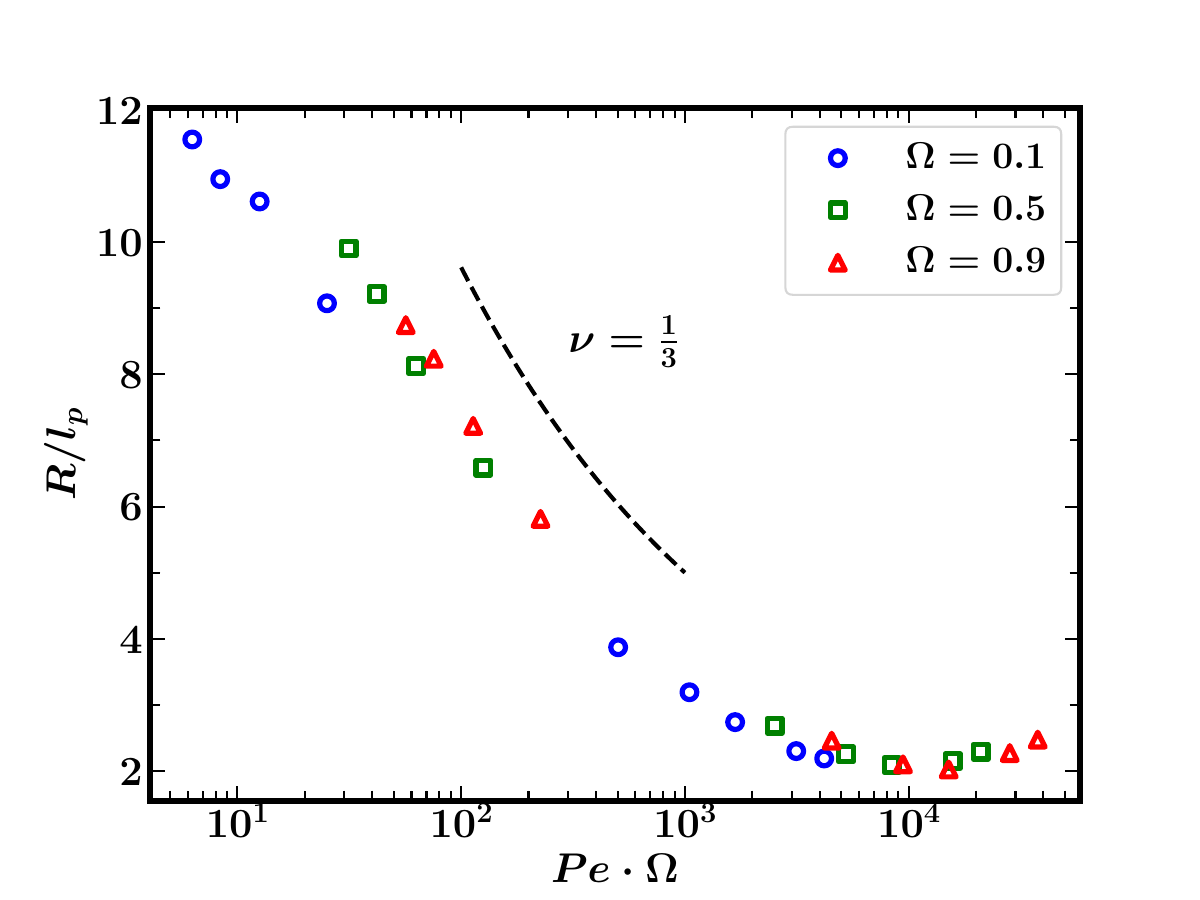}
\caption{Scaling of the normalized end-to-end distance $R/l_p$ with effective activity ${\rm Pe}\, \Omega$ for $u = L/l_p = 3.33$. The data collapse across three values of bare processivity ($\Omega = 0.1$, 0.5, 0.9) indicates a robust scaling. The dashed line plots $R/l_p \sim ({\rm Pe} \cdot \Omega)^{-\nu}$, with exponent $\nu = 1/3$.
}
\label{scaling}
\end{figure}
Spiral tightness increases with activity, as reflected by higher turning numbers and reduced end-to-end distance $R$, which collapses onto a single curve when plotted against the effective activity ${\rm Pe} \times \Omega$ (Fig.~\ref{scaling}).

A mean-field estimate captures this trend by balancing the active torque $f_a R^2$ with the bending moment $k_B T l_p / R$, yielding:
\begin{equation}
    R \sim \left(\frac{k_B T l_p}{f_a}\right)^{1/3} \nonumber
\end{equation}

Assuming $f_a \sim {\rm Pe} \, \Omega$, this gives the scaling:
\begin{equation}
    R \sim ({\rm Pe} \, \Omega)^{-1/3}
\end{equation}

Figure~\ref{scaling} shows good agreement with this prediction across a range of ${\rm Pe} \, \Omega$.

\begin{figure}[htbp]
\centering
\includegraphics[width=0.7\columnwidth]{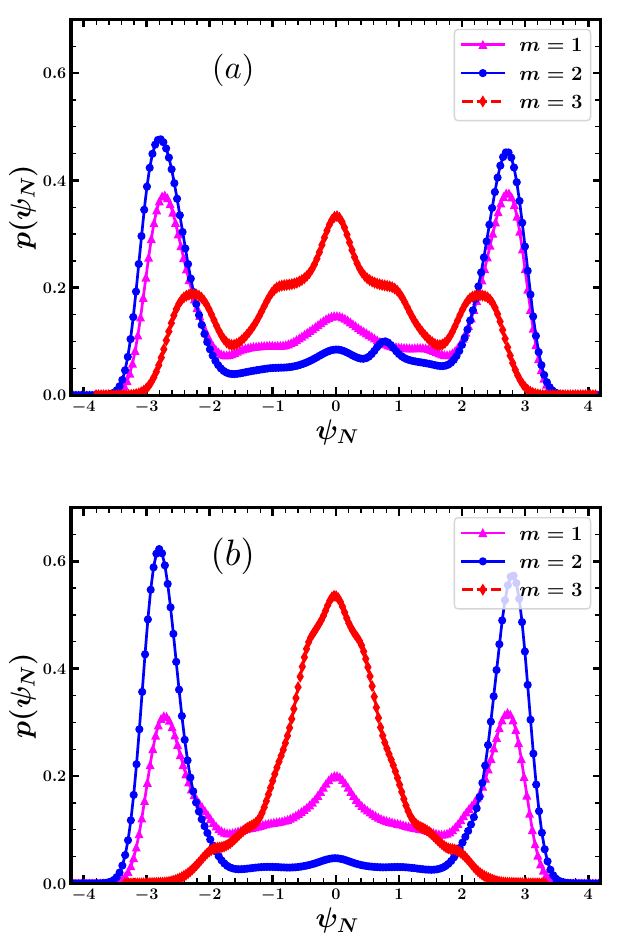}
\caption{Probability distribution of the turning number ($\psi_N$) for a polymer of length $L = 63\sigma$, with (a) $N = 96$  and (b) $N = 64$.}
\label{discrete}
\end{figure}
\section{Effect of Polymer Discretization on Spiral Stability}
\label{appendix:discretization}
In the main text, we observe that spirals are more stable for $m = 2$ compared to $m = 1$. Note that even for a free polymer driven by the motility assay, the force distribution at the end bead is inherently asymmetric, since motor forces act only on one side of the filament, unlike in the bulk, where forces are more evenly distributed across bonds. Consequently, the end bead itself acts as an effective defect, enhancing local fluctuations towards the formation of spirals.

To examine the influence of discretization on spiral stability, we increased the number of beads from $N = 64$ to $N = 96$ while keeping the polymer length $L$ fixed. With this finer discretization, the segment length  $r_0$ decreases, resulting in a smoother distribution of forces and curvature along the filament. As a result, the disparity in spiral stability between $m = 1$ and $m = 2$ diminishes, with spirals initiated with $m = 1$ pivot becoming more stable, while those at $m = 2$ pivot exhibit reduced stability. Furthermore, due to reduced spacing for finer discretization of the polymer, the pivot at $m=3$ position provides stronger spiral stabilization, as is evident from a comparison between Fig.~\ref{discrete}(b) and (a).

\section{Kurtosis of turning-number distributions}
\label{appendix:kurtosis}
To complement the phase diagrams in Sec.~\ref{sec:phase_diagram}, we plot the kurtosis ${\cal K}$ of $\psi_N$ for various parameter sets.

\begin{figure}[t]
    \centering
\includegraphics[width=0.8\columnwidth]{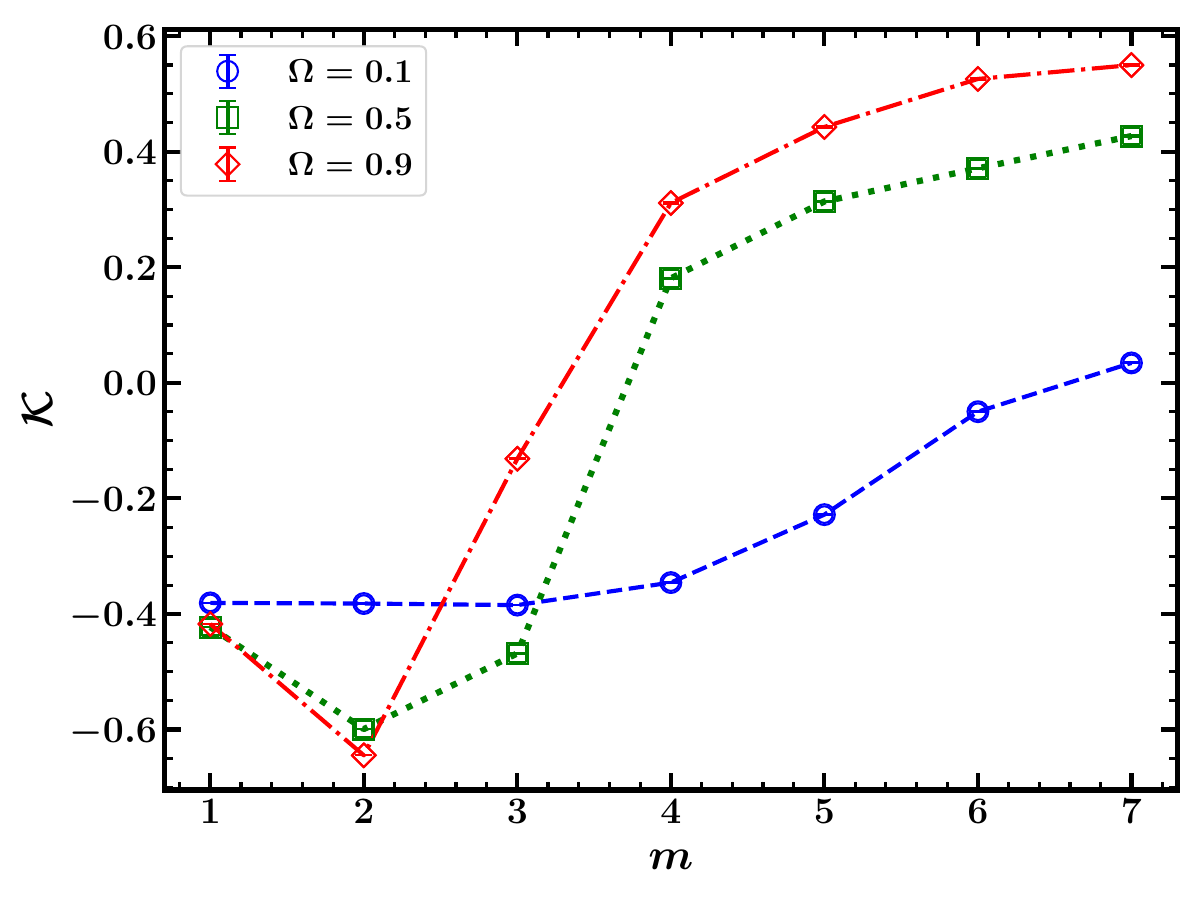}
    \caption{Kurtosis $\mathcal{K}$ of $\psi_N$ as a function of pivot position $m$ 
    for $\text{Pe}=10^5$ and processivity $\Omega=0.1,\,0.5,\,0.9$.}
    \label{fig:kurtosis_vs_m}
\end{figure}

Figure~\ref{fig:kurtosis_vs_m} shows ${\cal K}$ versus pivot position $m$ at fixed $\text{Pe} = 10^5$ and $\Omega = 0.1,\,0.5,\,0.9$. Near the polymer end, ${\cal K}$ is negative  -- approaching the spiral-signature value ${\cal K} = -2/3$ at $m=2$ -- indicating most stable spiral formation. As $m$ increases, ${\cal K}$ approaches zero or becomes positive, signaling a transition to open or flagellar beating states. This transition occurs at larger $m$ for low $\Omega$ and at smaller $m$ for higher $\Omega$.

\begin{figure}[t]
    \centering
\includegraphics[width=0.8\columnwidth]{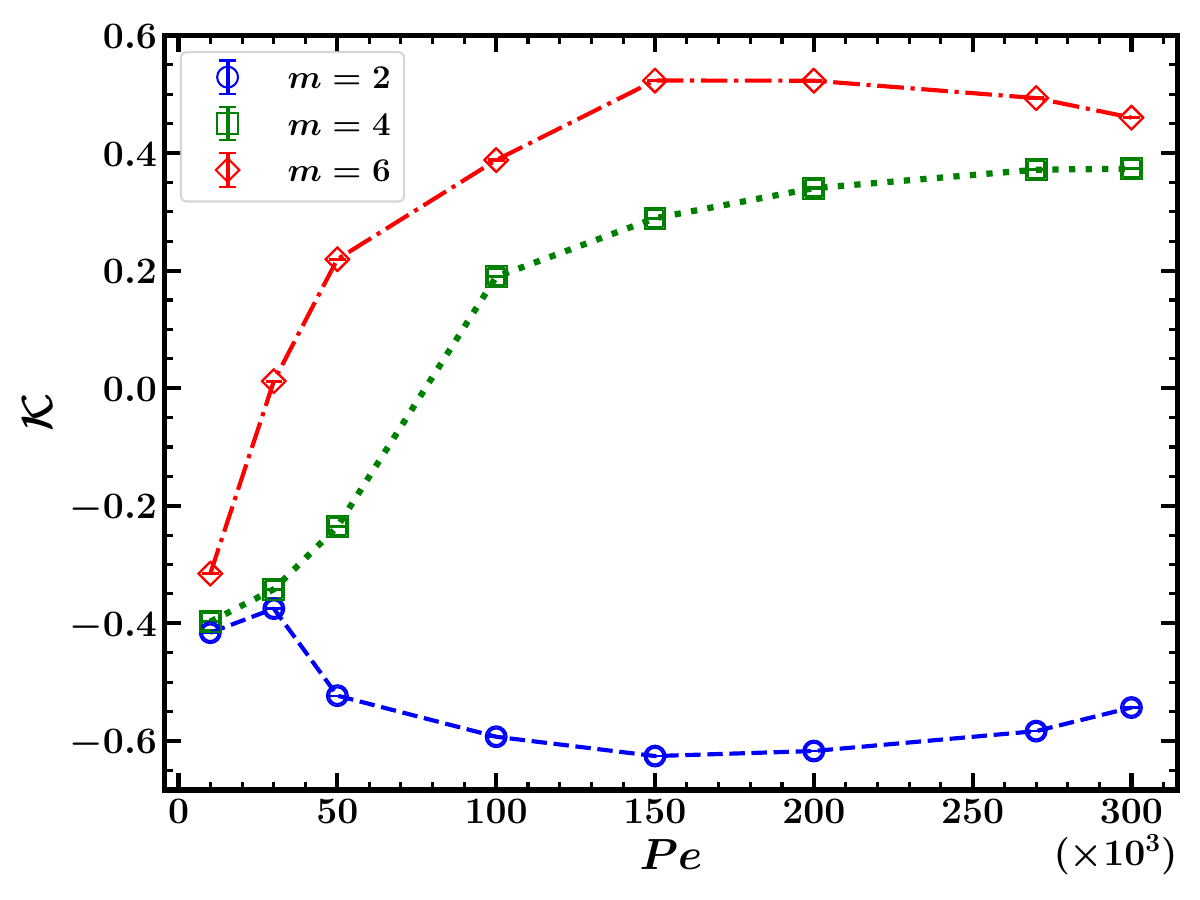}
    \caption{Kurtosis $\mathcal{K}$ of $\psi_N$ as a function of $\text{Pe}$ for fixed 
    $\Omega = 0.5$ and pivot positions $m=2,\,4,\,6$. }
    \label{fig:kurtosis_vs_pe}
\end{figure}

Figure~\ref{fig:kurtosis_vs_pe} plots ${\cal K}$ versus $\text{Pe}$ for fixed $\Omega = 0.5$ and $m=2,\,4,\,6$. At small $m$, ${\cal K}$ becomes more negative with increasing $\text{Pe}$, indicating stronger spiral stability, before reversing at high $\text{Pe}$ due to increased motor detachment. For larger $m$, ${\cal K}$ stays near zero or positive across all $\text{Pe}$, consistent with suppressed spiral formation.

\begin{figure}[htbp]
    \centering
\includegraphics[width=0.48\textwidth]{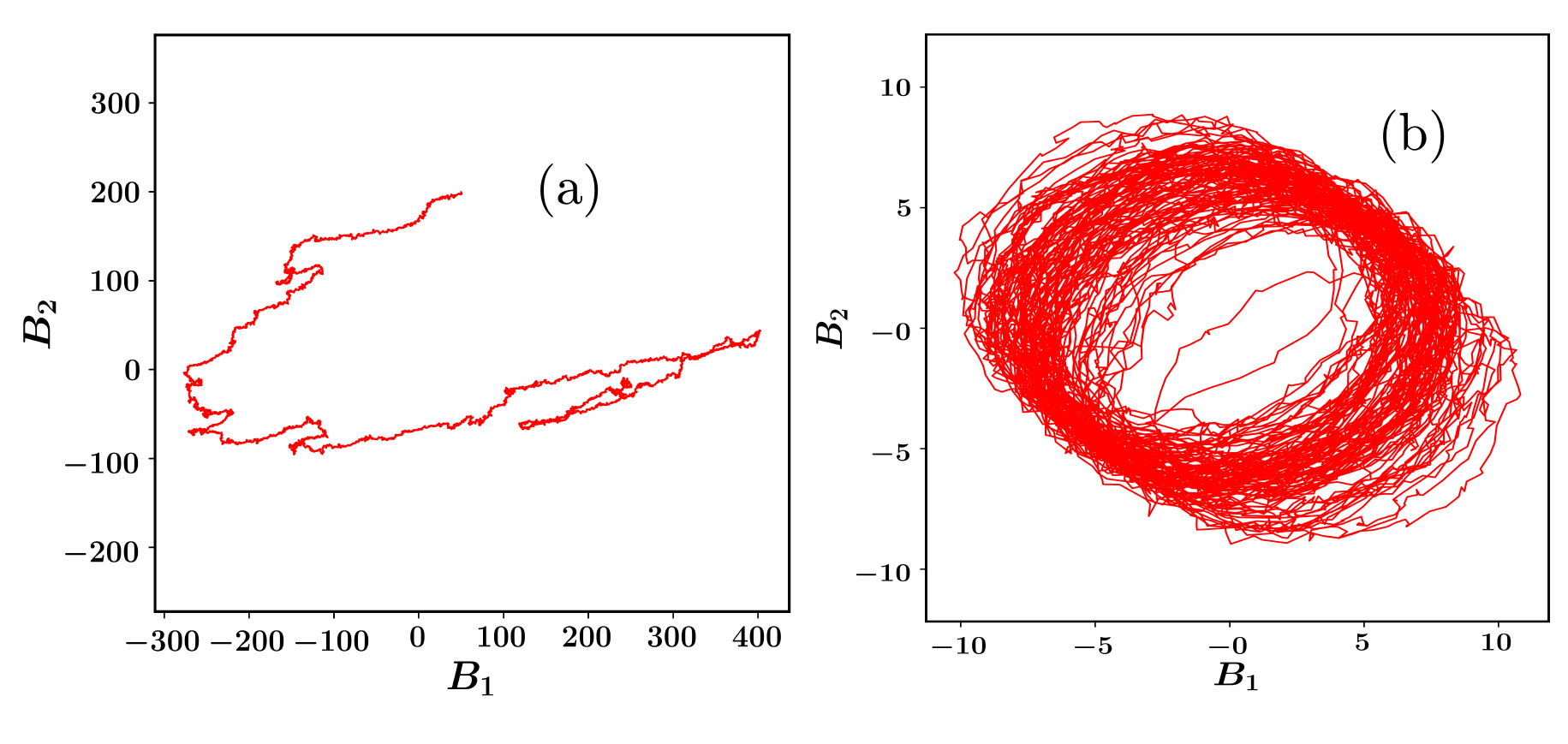}
\caption{
Parametric plots of the first two PCA mode amplitudes $(B_1(t), B_2(t))$ for (a) $m = 1$ and (b) $m = 40$, at $\text{Pe} = 1.78 \times 10^5$ and $\Omega = 0.5$. Panel (b) illustrates a stochastic limit cycle characteristic of flagella-like beating.
}
    \label{fig:limit}
\end{figure}

\begin{figure}[htbp]
\centering
\includegraphics[width=0.8\columnwidth]{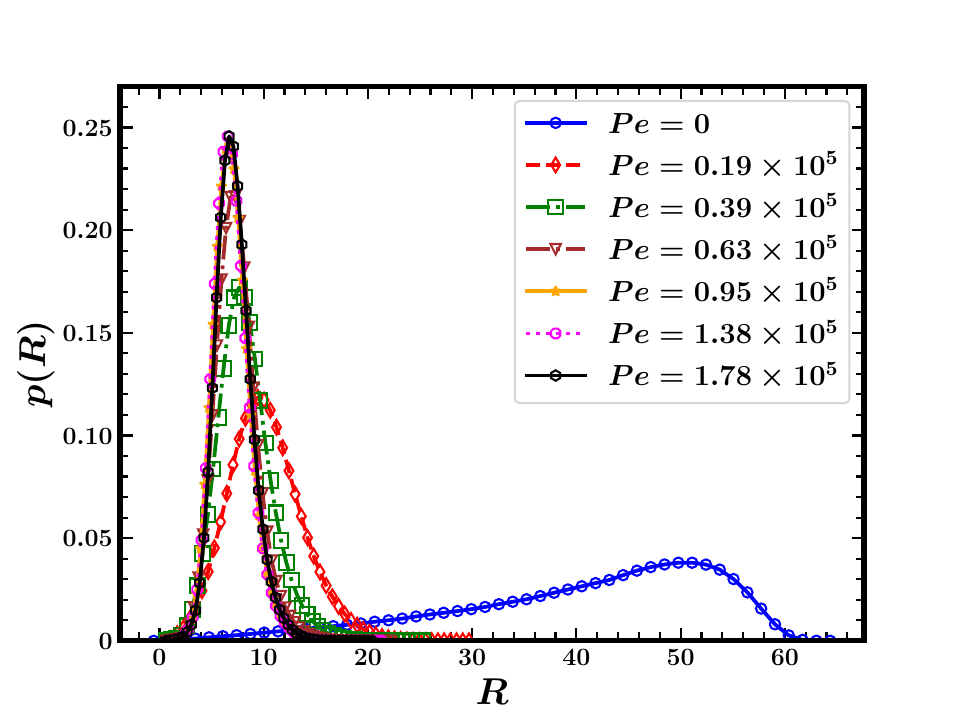}
\caption{Probability distribution of the end-to-end distance with varying $\text{Pe}$ at $\Omega = 0.5$ }
\label{fig:pend}
\end{figure}

\section{PCA analysis}
\label{appendix:pca}
 In the PCA analysis, we considered the tangent angle 
$\Psi = \psi[i,t]$ at each monomer, storing it for every time 
instant to form a data matrix~\cite{ma2014active}. After subtracting the mean tangent 
angle $\Psi_0$ from each entry, we constructed the symmetric 
covariance matrix
\[
\mathcal{C} = (\Psi - \Psi_0)\cdot (\Psi - \Psi_0)^{T}.
\]
The eigenvectors of $\mathcal{C}$ define the shape modes, 
denoted $X_1, X_2, \dots$. Any instantaneous polymer shape 
can be expressed as
\[
\psi(s,t) = \psi_0 + B_1(t) X_1(s) + B_2(t) X_2(s) + \dots
\]
For chains exhibiting flagella-like beating, the trajectory in the phase space of the first two mode amplitudes, $B_1(t)$ and $B_2(t)$, forms a closed limit cycle. In contrast, no such cycle appears when periodic beating is absent, as shown in Fig.~\ref{fig:limit}(a). Figure~\ref{fig:limit}(b) displays a broadened limit cycle, reflecting stochastic fluctuations in the filament dynamics. This cycle results from a combination of two motions: (i) swinging of the free end and (ii) oscillatory traveling waves along the backbone.

We compute the phase angle $\phi(t) = \tan^{-1} \left( {B_2(t)}/{B_1(t)} \right)$ and extract the mean angular velocity $\langle \dot{\phi} \rangle$ as a measure of beating frequency. To avoid cancellation from changes in rotation direction, we use the absolute value of $\dot{\phi}$ for the average. This frequency exhibits a pronounced minimum at $m = 2$, corresponding to tight spiral states, and increases with $m$, eventually saturating at larger pivot positions (Fig.~\ref{fig:transition}).

\section{End-to-end probability distribution}
\label{appendix:endtoend}
The end-to-end distance distribution $p(R)$ peaks at large $R$ for $\text{Pe} = 0$, reflecting extended conformations typical of self-avoiding semiflexible filaments (Fig.~\ref{fig:pend}). With increasing $\text{Pe}$, the polymer coils more tightly, leading to a sharper peak at lower $R$.


%

\end{document}